\documentstyle[11pt,paspconf,epsf]{article}
\def\etal{{et~al.~}}
\def\Lya{{\rm Ly}\kern 0.1em$\alpha$}
\def\HI{{\rm H}\kern 0.1em{\sc i}}
\def\Mg{{\rm Mg}\kern 0.1em{\sc ii}}
\def\Mgb{{\bf Mg}\kern 0.1em{\sc ii}}
\def\MgI{{\rm Mg}\kern 0.1em{\sc i}}
\def\Fe{{\rm Fe}\kern 0.1em{\sc ii}}
\def\Ca{{\rm Ca}\kern 0.1em{\sc ii}}
\def\MgII{{\rm Mg}\kern 0.1em{\sc ii}~$\lambda\lambda2976, 2803$}
\def\C{{\rm C}\kern 0.1em{\sc iv}}
\def\NV{{\rm N}\kern 0.1em{\sc v}}
\def\CIV{C\kern 0.1em{\sc iv}~$\lambda\lambda1548, 1550$}
\def\kms{\hbox{km~s$^{-1}$}}
\def\cm2{\hbox{cm$^{-2}$}}

\begin{document}

\title{{\Mgb} Absorbers: Disks, Halos, Satellites, and Pairs?}

\author{Jane C. Charlton\altaffilmark{1} and 
       Christopher W. Churchill\altaffilmark{2}}
\affil{Astronomy and Astrophysics Department, Pennsylvania State University,
       University Park, PA 16802}

\altaffiltext{1}{Center for Gravitational Physics and Geometry,
             Pennsylvania State University}
\altaffiltext{2}{Board of Studies in Astronomy and Astrophysics,
       University of California, Santa Cruz, CA 95064}

\begin{abstract}
To understand which parts of galaxies give rise to the variety of
observed {\Mg} absorption profiles seen in QSO spectra, we have
embarked upon a program to generate simulated absorption profiles as
they would arise from the many typical galactic structures we see
today.  Here, we present preliminary results for a clumpy, rotating
disk which has been sampled by a variety of QSO lines of sight.
The resulting ensemble of simulated profiles is qualitatively similar
to those observed in HIRES spectra of  $0.4 < z < 1.0$ {\Mg} systems
(Churchill, this volume).  Additionally, we discuss the statistical
contribution from satellite galaxies, which would be expected to
contribute additional absorption subcomponents and are likely to have
been more numerous at intermediate redshifts.  Our preliminary finding
is that the common disk and satellite components of galaxies may, to a
large degree, be the structures giving rise to a variety of the
observed {\Mg} absorption profiles.
\end{abstract}


\section{Introduction}

As seen in spectra from HIRES (Vogt \etal 1994) on Keck, the {\Mg}
absorption profiles arising in $0.4 < z < 1.0$ galaxies exhibit a rich
variety of subcomponent structure and kinematic complexity (Churchill,
Steidel, \& Vogt 1996; Churchill, this volume).
The key to applying QSO absorption line studies to the
problem of galaxy evolution hinges upon establishing the connections
between absorption--line properties and the components and processes
in galaxies that give rise to the absorption.
One simple absorption property is the ``morphological
or kinematic shape'' of the profiles, which we explore here.

The status--quo picture of {\Mg} absorbers is that absorption
arises from ``clouds'' spatially distributed throughout galactic
halos.  
The covering factor of these clouds is inferred to be nearly
unity, since intermediate redshift galaxies that lie within 35--40~kpc
of a QSO line of sight rarely fail to exhibit {\Mg} absorption
[Steidel 1995 (S95)].
However, in the present--day Milky Way, only a $14 \pm 8$\% covering
factor is observed for {\Mg} down to  $W_0 \sim 0.05$~{\AA} (Bowen,
Blades, \& Pettini 1995).
The implication is that either halo clouds may not be the {\it
primary}\/ source of {\Mg} absorption or that the spatial distribution
of halo gas has evolved substantially from $z \sim 0.4$ to the
present.

We are exploring the theme: ``what you see is what you get''.   
In other words, we examine the hypothesis that the galactic substructures
known from local galaxies are sufficient to give rise to the observed
variety of {\Mg} absorption profile morphologies.  Our first goal is
to ascertain if models based upon this hypothesis fail to produce
simulated profiles similar to those observed.
The two components {\it known}\/ to exist in local
galaxies that we explore here are:
\begin{enumerate}
\item Extended and warped {\HI} disks of spiral/disk galaxies,
\item Dwarf satellite galaxies associated with typical
$L^{\ast}$ galaxies.
\end{enumerate}
{\Mg} is associated with {\HI} down to a neutral column density
$N_H \sim 10^{17}$~{\cm2}, so one issue is how extended
these disks are to this {\HI} column density.
Likewise, we need to estimate the gaseous extent of dwarf satellite
galaxies and the number of dwarfs that are associated with a typical
$L^{\ast}$ galaxy.
We have assumed that no dramatic evolution in the spatial distribution
of these absorbing structures has occurred from $z \sim 0.4$ to the
present and have preliminarily explored whether these present--day and
familiar components can qualitatively reproduce the morphological
shapes and variety of observed {\Mg} absorption profiles.

\section{Contribution to {\Mgb} Absorption From Galaxy Disks}

Naively, one would at first expect that highly inclined disk galaxies,
observed over the full range of impact parameters, would not
frequently give rise to {\Mg} absorption.  
However, when we account for the increased pathlength and
velocity dispersion sampled through an inclined disk, we find that
``clumpy disk'' models yield nearly the same absorption properties as
do models of ``clouds'' distributed in a spherical halo [Charlton \&
Churchill 1996 (CC96)].
Both models yield an effective covering factor of $\sim 80$\%, and so
are in apparent conflict with the very small number of non--absorbing
galaxies (only 3/51) reported to lie below the $D = 38h^{-1}
(L_K/L_K^*)^{0.15}$~kpc ``absorption boundary'' (S95).
However, the survey work of Steidel, Dickinson, \& Persson (1994)
reported in S95 is not fully complete.  
Once their observational selection procedures to date are considered,
and the possibility of an occasional absorber misidentification is
accounted for, the discrepancy in covering factor between our models
and the data may be resolved (CC96).
Thus, both spherical or disk models are currently consistent with the
available data.
We have developed simple tests useful for distinguishing between these
two geometries.  
These tests rely upon high--spatial resolution imaging of the galaxies
and high--resolution spectra of the absorption (CC96). 

We now address ``what you get'' in {\Mg} absorption when you sample
lines of sight through a galaxy disk of various orientations.  
We designed a Monte Carlo clumpy disk with empirical properties
roughly consistent with those seen in nearby spiral galaxies.  
We randomly distributed ``absorbing clouds'' in a disk, which we have
assumed to be increasing in thickness from 1~kpc to 10~kpc and to have
a truncation radius of 50~kpc. 
The individual clouds have radii of 1~kpc and number densities given by
$n_{cl} \sim 1/R$ (so that the equivalent width $W \sim 1/R$).
The cloud--cloud velocity dispersions were $\sim 6$~{\kms}.
These models were viewed from random orientations at a given impact
parameter, and the resulting absorption profiles were sampled every
2~{\kms}.
The profiles were then convolved with the HIRES instrumental
profile, and Poisson noise was added to give a $S/N \sim
30$.  In Figure 1, we show a series of simulated {\Mg} profiles with
increasing impact parameter, which allows qualitative comparison with
observed HIRES spectra (Fig.~1 of Churchill, this volume).

\begin{figure}[th]
\plotone{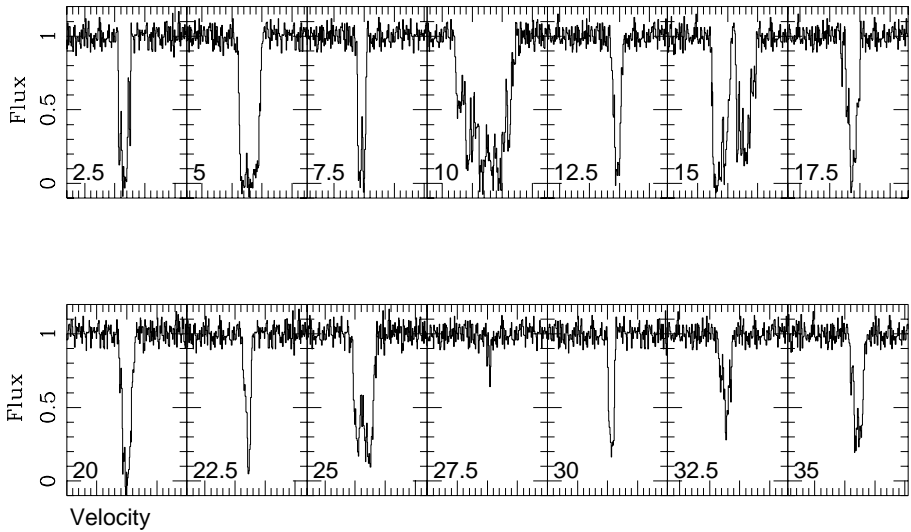}
\vskip -2.3in
\caption{
Simulated {\Mg} absorption profiles obtained by viewing a clumpy,
rotating disk at random orientations.  The impact parameter is
increased in increments of 2.5~kpc through the series.  Each panel has
a velocity spread of 400~{\kms}.}
\end{figure}

We have also found that a series of model disks viewed with increasing
inclination (at fixed impact parameter) show an increasing spread in
subcomponent velocity and a gradual progression toward greater
kinematic complexity (i.e.~blended components separate in velocity
space).
Thus, simulations such as those presented here will yield an
orientation test, once high--spatial resolution images of the
absorbing galaxies are available.

\section{Contribution to {\Mgb} Absorption From Satellite Galaxies}

A major qualitative difference between the simulated absorption
spectra from the clumpy disk models and the observed profiles is the
presence of ``high--velocity'' weak features in the observed data.
We consider the probability that lines of sight through $L^{\ast}$
galaxies may also pass through dwarf satellites, which would give rise
to outlying and relatively weak lines.
Assuming that the gaseous cross section of a satellite is given by
$R_{sat} = 35 (L/L_{B*})^{0.2}$~kpc, we computed the mean number of
satellites intercepted by random lines of sight at various impact
parameters through a Milky--Way like galaxy in a ``Local Group''
(using the actual distances and luminosities of Local Group members).
The overall probability of intercepting a satellite ranges from 0.2
to 0.35, with the dominant contribution being LMC and SMC--like
satellites.  

How many satellites would be needed to produce a unity probability of
interception?  
Assuming that the luminosity distribution of dwarf satellites can be
described by a Schechter function with $\alpha = -1$ ($-18.5 < M_B <
-12.5$), and that they follow an isothermal spatial distribution
between $0 < R < 400$~kpc around the primary, we find that that $\sim
20$ satellites are required.
There are roughly 4--5 dwarf satellites associated with the Milky Way
that fit these specifications.
Thus, evolution in the numbers of dwarf galaxies associated with
$L^{\ast}$ galaxies would be required if these ``high--velocity'' weak
features do arise from dwarf satellites.
If we extrapolate the results of Carlberg, Pritchett, \& Infante
(1995) that the number of close pairs increases with redshift 
as $(1+z)^{3.4}$, we find 25--30 satellites would be expected per
Milky Way by $z=0.7$.  
If so, an even larger number of ``high--velocity'' {\Mg} absorption
lines would be predicted from $z > 1$ galaxies.

\section{Is What You See What You Get?}

A simple preliminary clumpy disk model appears to produce simulated
{\Mg} profiles remarkably similar to those observed from $0.4 < z <
1.0$ galaxies.
Combined with a plausible contribution from satellite galaxies, 
we might find that these simulated profiles would also exhibit
the higher--velocity weaker features commonly observed.
We strongly caution, however, that this preliminary model is far from
unique.  
Our future efforts will include modeling distributions of clouds in a
spherical geometry with various velocity fields (eg., rotation, radial
infall, and isotropic).
If the simulated spectra from familiar local galactic structures and 
systematic velocity fields match the ``morphological'' and kinematic
substructures of observed profiles, then there may not be so much more
behind producing {\Mg} absorption than what meets the eye.
It would be fruitful to observationally confirm if higher--velocity
weaker features are actually produced by satellite galaxies and to
establish if the orientation effects expected from disks are seen.
Such a study would depend upon deep high--spatial resolution images
(roughly 20--30) of the galaxies for which HIRES spectra are
available.

\acknowledgments
We acknowledge NASA grant NAGW--3571.

\end{document}